\pdfoutput=1
\documentclass[reprint,amsmath,amssymb,aps,prresearch,superscriptaddress,floatfix]{revtex4-2}

\usepackage{graphicx}
\usepackage{dcolumn}
\usepackage{bm}
\usepackage{hyperref}

\begin{document}

\title{Efficient Ground-State Manifold Amplification for \\ Rydberg Atom Arrays via Houdayer Cluster Updates}

\author{Andrew J.~Ochoa}
\affiliation{JIJ Inc., Minato-ku Tokyo, 108-0023 Japan}
\affiliation{Strangeworks Inc., Austin, Texas, 78702, USA}

\author{Martin J.~A.~Schuetz}
\affiliation{Amazon Advanced Solutions Lab, Seattle, Washington 98170, USA}
\affiliation{AWS Quantum Technologies, Seattle, Washington 98170, USA}

\author{Helmut G.~Katzgraber}
\affiliation{55 North Management ApS, Ole Maaløes Vej 3, 2200 København, Denmark}  

\date{\today}

\begin{abstract}

We demonstrate a hybrid quantum-classical approach for exploring
ground-state manifolds of maximum independent set problems, in which
quantum hardware supplies the initial solutions and an entirely
classical post-processing algorithm expands them, increasing the
solution diversity of quantum hardware measurements by several orders of
magnitude. Quantum optimization devices, while capable of generating
high-quality solutions, can suffer from fundamental sampling limitations.
This is particularly problematic for applications that strongly depend
on a large variety of uncorrelated optimal solutions. We apply the
Houdayer cluster update algorithm to systematically expand
quantum-generated solution sets through classical post-processing. We
validate our approach on a test bed of $43$ problem instances from
QuEra's Aquila analog annealing device, achieving diversity
amplification factors, defined as the increase in the number of distinct
solutions at a given set size, of up to $11923\times$ with a median of
$24.3\times$ across different problem scales and hardness parameters.
Because exact ground-state degeneracies are known for these instances,
the recovered fraction of the manifold can be measured rather than
bounded: post-processing recovers a median of $21$\% of the ground-state
manifold, and provably all of it for six instances. We also demonstrate
that, in some cases, the classical post-processing can discover
independent sets larger than those found by the quantum hardware, thus
improving on solution quality and not just diversity. While the method
is non-ergodic and cannot guarantee complete manifold coverage, it
consistently discovers a larger number of high-quality, and in some
cases strictly better, configurations than quantum sampling alone. This
work demonstrates how simple classical post-processing can enhance the
utility of quantum optimization results.

\end{abstract}

\maketitle

\section{Introduction}
\label{sec:introduction}

The maximum independent set (MIS) problem is a paradigmatic NP-hard combinatorial optimization problem~\cite{karp1972}, which remains NP-hard on unit disk graphs~\cite{clark1990}, with applications spanning network design, computational biology, and portfolio optimization. Given an undirected graph $G = (V, E)$, the goal is to find the largest subset $S \subseteq V$ such that no two vertices in $S$ are adjacent. Neutral atom arrays with programmable geometries have emerged as a leading platform for quantum optimization of MIS problems~\cite{pichler2018,ebadi2022,serret2020}. In these systems, atoms are positioned with sub-micrometer precision and coupled via Rydberg blockade interactions that naturally encode the MIS problem on unit disk graphs, as originally proposed in Ref.~\cite{pichler2018}. The blockade radius $r_b$ defines the graph connectivity: edges connect atoms within distance $r_b$, while independence constraints arise from the strong energetic suppression, rather than strict impossibility, of simultaneous Rydberg excitation within the blockade sphere. Recent experiments on Rydberg atom arrays~\cite{ebadi2022}, including QuEra's Aquila quantum device~\cite{quera2023aquila,qredumis2025}, have demonstrated quantum annealing algorithms (QAA) capable of generating high-quality solutions of MIS problems with hundreds of atoms.

However, quantum optimization devices face a fundamental sampling limitation. While capable of accessing optimal or near-optimal configurations, quantum measurements from a typical hardware run produce only tens to hundreds of distinct (degenerate) solutions (for example, ranging from $2$ to $493$ in the data set of Ref.~\cite{qredumis2025} studied here). Moreover, quantum annealers have been shown to be biased samplers, both on superconducting hardware~\cite{mandra:17,konz2019fairsampling} and on Rydberg atom arrays~\cite{schuetz2025compilation}: they do not find all ground states with equal probability. This biased sampling becomes problematic when the ground-state manifold contains exponentially many degenerate optimal configurations as is often the case for the MIS. However, the classical post-processing presented in this work reveals that the aforementioned unit disk graph instances can have as many as $10^3$ to $10^6$ distinct maximum independent sets, with the largest manifold explored containing nearly 2 million optimal configurations. These counts are specific to instances of a few hundred atoms, and the degeneracy grows rapidly with size, so larger devices should make the sampling problem worse rather than better. The limited sampling diversity of quantum hardware leaves vast regions of these degenerate manifolds unexplored, restricting practical utility for applications requiring multiple diverse solutions such as ensemble methods and entropy calculations, design space exploration, post-optimization inclusion of constraints, or robustness analysis, to name a few. 

Classical post-processing provides a mechanism to amplify quantum-generated solution diversity. In previous work~\cite{feeding2019}, we introduced an algorithm based on Houdayer cluster updates~\cite{houdayer2001} to improve sampling from biased optimizers. The approach operates through rejection-free cluster exchanges at zero temperature: given two independent sets $\{s_i\}$ and $\{s_j\}$, the algorithm computes their symmetric difference $\Delta_{ij} = s_i \oplus s_j$ (vertices present in exactly one solution), identifies the $c$ connected components (clusters) in the induced subgraph $G[\Delta_{ij}]$, which we refer to as the disagreement graph, and generates $2^{c}$ candidate configurations by independently choosing which clusters to exchange. This mechanism preserves solution quality, and can occasionally improve it (Section~\ref{sec:classical_improves_quantum}), while enabling large-scale structural rearrangements that can reach configurations unavailable to local moves.

While fundamentally non-ergodic, i.e., offering no guarantee of complete manifold coverage, the method proved effective at reducing sampling bias in quantum annealing data from D-Wave systems and improving diversity for classical spin-glass solvers~\cite{feeding2019}, whose sampling biases are documented in Refs.~\cite{moreno:03,matsuda:09,mandra:17,konz2019fairsampling}.

The present work applies this approach to MIS measurement records from Rydberg atom arrays. Applied to two independent sets, the cluster exchange keeps every candidate independent by construction (Section~\ref{sec:houdayer_algo}), so none is rejected as infeasible. We use the method to ask how much ground-state information a neutral atom measurement record contains beyond the solutions the device reports. Because the instances studied here come with exact reference optima and exact ground-state degeneracies, we can measure both how many additional optimal configurations the record yields and what fraction of the true manifold that represents. We source initial solution pools from quantum hardware runs on QuEra's Aquila quantum device, using quantum measurements as high-quality seeds for classical cluster-based exploration. Raw hardware measurements may select adjacent atoms, and these are repaired through a two-stage greedy procedure~\cite{qredumis2025}. The first stage iteratively removes the vertices with most conflicts until independence is restored, and the second greedily adds vertices that maintain independence to recover maximality. This repair process runs in $\mathcal{O}(|E|)$ time per candidate and typically preserves optimal or near-optimal set sizes.

We evaluate our approach on $43$ problem instances across three categories whose hardness parameter spans several orders of magnitude: Benchmark instances ($137$ atoms each), hard instances~\cite{qredumis2025} that classical kernelization alone cannot reduce to an empty kernel ($115$ -- $231$ atoms), and standard instances ($7$ -- $205$ atoms). The quantum seeds come from measurements on QuEra's Rydberg quantum device, with solution counts ranging from $2$ to $493$ distinct configurations per instance at the size the hardware reached. Our classical post-processing systematically processes quantum solution pairs through Houdayer cluster updates, applying adaptive stopping criteria to balance manifold exploration against computational cost.

The results demonstrate amplification across the full dataset: solution diversity increases by factors ranging from $1\times$ to $11923\times$, with the quantum seed pool expanded to contain hundreds or thousands of distinct configurations at the seed size in seconds to minutes on commodity hardware \cite{hardware}. Discovery dynamics reveal that manifold exploration proceeds efficiently until saturation, with most new solutions discovered within the first few hundred to few thousand pair processings. We develop stopping criteria based on discovery rate saturation (halt when the number of newly discovered solutions over a sliding window of $500$ consecutive pairs falls below $0.1$\% of the current pool size) and computational budgets ($10^5$ pair limit), enabling practical termination without exhaustive enumeration.

For three instances (\texttt{L12\_seed829}, \texttt{L17\_seed424}, and \texttt{L17\_seed46}), the classical algorithm discovered independent sets \emph{larger} than those found by quantum hardware yielding solution quality improvements, not merely diversity amplification within a known manifold. These cases involve size increases from $35$ to $36$ atoms (\texttt{L12\_seed829}) and from $68$ and $67$ to $70$ atoms (\texttt{L17} instances), demonstrating that classical post-processing can improve upon quantum optimization results for analog quantum optimizers that are inherently noisy. As we detail below, these improvements reach the exact optimum in one case and come within one atom of it in the other two.

This work makes the following contributions:
\begin{enumerate}
\item \textbf{How much a measurement record contains}: quantification, across $43$ instances spanning three hardness families, of the additional distinct configurations recoverable from a neutral atom measurement record beyond those the device reports, a median of $24.3\times$ overall and $8.9\times$ over the $35$ instances whose seeds were already optimal, reaching $11923\times$ for \texttt{HP1478}, for $18$ minutes of laptop time across the entire dataset, measured end to end.
\item \textbf{Coverage measured against exact degeneracies}: because exact ground-state degeneracies are available for these instances, the recovered \emph{fraction} of the manifold is measured rather than bounded, a median of $21$\% and provably complete on six instances. This converts the non-ergodicity of the method from a disclaimer into a measured quantity, and shows that completeness is attainable precisely when the manifold is small enough for a fixed seed budget to enumerate.
\item \textbf{An audit of what the hardware determined}: comparison against exact reference optima shows the quantum seed pool sat strictly below the true optimum on $8$ of the $43$ instances, including six of the nine hard instances. On three of these, classical recombination discovered strictly larger independent sets than the hardware found, reaching the exact optimum on one. Distinguishing an amplified true manifold from an amplified suboptimal one requires such a reference.
\item \textbf{Adaptive stopping criterion}: saturation-based termination that keeps the recovery cost negligible relative to the hardware run being post-processed, with a bounded worst case.
\item \textbf{Application of cluster updates to Rydberg atom array MIS}: We
apply the cluster updates previously developed for classical solvers and superconducting annealers on quadratic binary optimization problems without field terms \cite{feeding2019}, to independent sets sampled by neutral atom hardware, with a feasibility argument showing that the exchange respects the independence constraint by construction.
\end{enumerate}

The remainder of this paper is organized as follows. Section~\ref{sec:methods} describes the instances studied, quantum hardware setup, Houdayer cluster update algorithm, MIS repair procedure, and the stopping criterion. Section~\ref{sec:results} presents amplification results, classical improvements, discovery dynamics, and efficiency analysis. Section~\ref{sec:discussion} discusses implications, limitations, and connections to related work. Section~\ref{sec:conclusion} concludes with future directions.

\section{Methods}
\label{sec:methods}

\begin{figure*}
\centering
\includegraphics[width=0.95\textwidth]{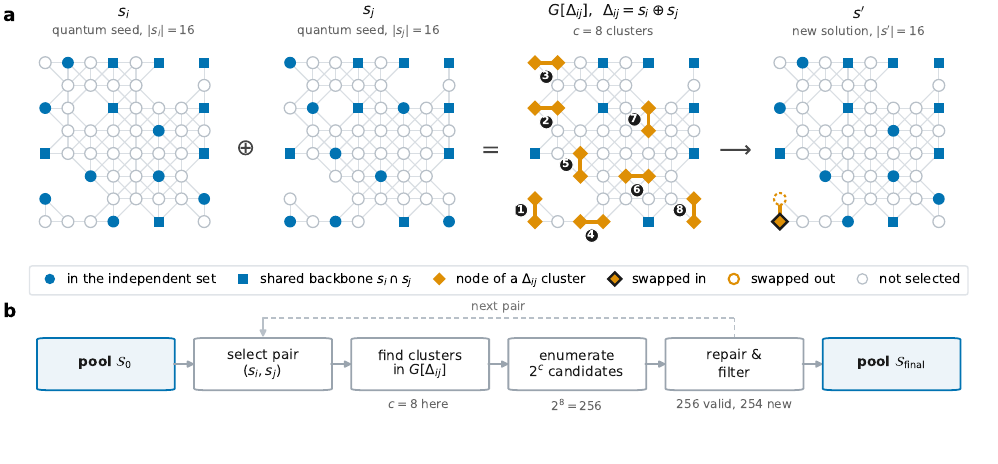}
\caption{Houdayer cluster update mechanism for ground-state manifold amplification. \textbf{(a)} A single update, shown on a real instance from our data set (\texttt{atoms\_L8\_seed3066}: $51$ atoms, $135$ unit-disk edges, $|\mathrm{MIS}| = 16$) and read left to right as an identity. Two solutions $s_i$ and $s_j$ returned by the QuEra Aquila hardware, both of optimum size $16$, are combined into their symmetric difference $\Delta_{ij} = s_i \oplus s_j$. Atoms belonging to a solution are drawn as filled blue markers and unselected atoms as open circles. The $8$ atoms selected in both solutions form the shared backbone (blue squares) and are untouched by the update; the $16$ atoms of $\Delta_{ij}$ (orange diamonds) split into $c = 8$ disjoint connected components of $G[\Delta_{ij}]$, numbered and joined by bars. Exchanging the atoms of any one cluster between the two solutions yields a distinct independent set of the same size; the fourth lattice shows the result $s'$ for cluster $1$, where one atom is swapped in (bold diamond) and one swapped out (dotted outline). Because the clusters are disjoint and the backbone is preserved, each of the $c$ clusters can be exchanged independently, so this single pair generates $2^c = 256$ candidate configurations. For this pair all $256$ are independent sets of size $16$, the maximum for this instance, and $254$ of them are new, illustrating why no Metropolis-style rejection step is required. \textbf{(b)} Algorithm flow, annotated beneath each step with the corresponding numbers for the pair in (a). Quantum measurements provide the seed pool $\mathcal{S}_0$, of size $N_{\mathrm{quantum}}$; the algorithm iterates over pairs drawn from that pool, enumerating and repairing $2^c$ candidates per pair. Iteration continues until saturation (newly discovered solutions over a $500$-pair sliding window fall below $0.1$\% of the current pool size), exhaustive pair processing, or a computational budget of $10^5$ pairs, yielding $N_{\mathrm{total}}$ solutions with amplification factor $A = N_{\mathrm{total}} / N_{\mathrm{quantum}}$.}
\label{fig:workflow}
\end{figure*}

Our approach follows the hybrid quantum-classical workflow illustrated in Figure~\ref{fig:workflow}. Quantum hardware runs on QuEra's Aquila quantum device generate an initial pool of candidate solutions for each problem instance. These quantum-generated samples are then passed to a classical post-processing pipeline based on Houdayer cluster updates~\cite{houdayer2001,feeding2019}, which systematically recombine solution pairs to discover further distinct configurations at the same set size. Whether that size is the true optimum depends on the seed pool and is established separately in Section~\ref{sec:validation}. The expanded solution pool is validated and the amplification factor is recorded. Throughout, $\mathcal{S}_0$ denotes the seed pool, the distinct repaired solutions at the largest size $k$ reached in the hardware record, and $\mathcal{S}_{\mathrm{final}}$ the solutions held at the final size $k_{\mathrm{final}} \geq k$ once post-processing terminates. The amplification factor is their ratio, $A = N_{\mathrm{total}} / N_{\mathrm{quantum}} = |\mathcal{S}_{\mathrm{final}}| / |\mathcal{S}_0|$, with $N_{\mathrm{quantum}} = |\mathcal{S}_0|$ and $N_{\mathrm{total}} = |\mathcal{S}_{\mathrm{final}}|$. For the three instances whose size improved, $A$ and all aggregate statistics use the first pass at the seed size, and their pools at the improved size are reported in Section~\ref{sec:classical_improves_quantum}. The following subsections describe each component of the pipeline.

\subsection{Problem Instances}
\label{sec:instances}

We evaluate our approach on $43$ problem instances sourced from quantum hardware runs on QuEra's Aquila quantum device from the study in Ref.~\cite{qredumis2025}, spanning three categories that cover a wide range of problem sizes and hardness regimes.

The \textbf{Benchmark Competition} category comprises four site-diluted union-jack graph instances (\texttt{HP125}, \texttt{HP1412}, \texttt{HP1435}, and \texttt{HP1478}), each with $137$ atoms on a square lattice of side length $L = 14$~\cite{qredumis2025}. The instance labels are mnemonics for the hardness parameter $H$ of Ref.~\cite{qredumis2025} rather than a literal digit-by-digit encoding of it (\texttt{HP1478}, \texttt{HP1412}, \texttt{HP125}, and \texttt{HP1435} denote $H = 1.478$, $14.12$, $125.5$, and $1435$, respectively). The hardness parameter is the classical, conductance-like degeneracy ratio $H = D_{|\mathrm{MIS}|-1} / (|\mathrm{MIS}|\, D_{|\mathrm{MIS}|})$, where $D_\alpha$ is the number of independent sets of size $\alpha$~\cite{andrist2023hardness}, and the QAA success probability is suppressed exponentially in $H$~\cite{qredumis2025}. These four values range from near-trivial to seemingly intractable for unassisted QAA. The reference data of Refs.~\cite{qredumis2025,qredumis_github} record $D_{|\mathrm{MIS}|-1}$ for the other $39$ instances, so $H$ is available for all $43$ and spans $1.03$ to $9717$.

The \textbf{Hard} category comprises nine instances ($115$ to $231$ atoms) on which Ref.~\cite{qredumis2025} ran its qReduMIS pipeline on hardware. Their seeds come from the full-graph QAA runs of the same reference, as for the other families (Section~\ref{sec:quantum_seeds}). On these instances the classical reduction rules stall before the kernel is eliminated, leaving a residual subproblem that quantum hardware is used to unblock, making them among the most challenging optimization problems in the dataset. 

The \textbf{Standard test bed} category comprises thirty $L = 3$ -- $L = 16$ instances ($n = L\times L$ lattice sites and with $7$ -- $205$ atoms) from the QAA hardware test bed~\cite{qredumis2025,ebadi2022}. These fully-reducible instances span a wide range of problem sizes and graph structures arising from the unit disk constraint. Full instance specifications and quantum measurement data for both categories are provided in Refs.~\cite{qredumis2025,qredumis_github}.

\subsection{Quantum Hardware and Solution Seeds}
\label{sec:quantum_seeds}

Each problem instance is encoded as a unit disk graph on QuEra's Aquila quantum device, where atoms are positioned at graph vertices and the Rydberg blockade radius defines graph connectivity~\cite{ebadi2022,quera2023aquila}. The device executes a time-dependent Hamiltonian evolution governed by global Rabi frequency and detuning controls, with the problem structure encoded through the atomic positions together with the blockade radius, which is itself set by the laser parameters. Following Ref.~\cite{qredumis2025}, with $\hbar = 1$, the Hamiltonian is
\begin{equation}
\begin{split}
\mathcal{H}(t) ={}& \frac{\Omega(t)}{2} \sum_{i} \left( e^{\mathrm{i}\phi(t)} |0\rangle_i \langle 1| + \mathrm{h.c.} \right) \\
&- \delta(t) \sum_{i} n_i + \sum_{i<j} V_{ij} \, n_i n_j ,
\end{split}
\label{eq:rydberg}
\end{equation}
where $|0\rangle_i$ and $|1\rangle_i$ are the ground and Rydberg states of atom $i$, $n_i = |1\rangle_i \langle 1|$ is its occupation operator, $\Omega(t)$ and $\phi(t)$ are the Rabi frequency and phase of the global drive, and $\delta(t)$ is the laser detuning. The interaction $V_{ij} = C_6/|\mathbf{x}_i - \mathbf{x}_j|^6$ is the van der Waals coupling between atoms at positions $\mathbf{x}_i$ and $\mathbf{x}_j$, with $C_6$ the van der Waals coefficient. Unlike the zero-field spin glasses for which Houdayer cluster updates were developed~\cite{houdayer2001}, this Hamiltonian contains a field term, the detuning $\delta(t)$ acting on every $n_i$. At the conclusion of each quantum anneal, whose duration is of order microseconds, projective site-resolved measurements yield a classical bit string indicating the Rydberg state of each atom. Raw outcomes that violate independence constraints due to finite blockade strength or imperfect adiabaticity are repaired using the procedure described in Section~\ref{sec:repair}. Repeating this process over many shots yields an initial pool $\mathcal{S}_0$ of distinct solutions at the largest size the hardware attained, which Section~\ref{sec:validation} compares against the certified optimum.

All $43$ seed pools derive from direct anneals on the full problem instance ($1000$ shots per instance~\cite{qredumis2025}, of which $49$ to $946$, median $426$, survive post-selection for correct atom loading), and the repair step leaves most of each measurement intact: on average $92$\% to $94$\% of the atoms in a repaired seed come from the measurement itself for the $30$ standard instances, and $90$\% to $93$\% for the nine hard instances. For the hard family these are the vanilla QAA runs on the full graph that Ref.~\cite{qredumis2025} reports as the baseline for its qReduMIS pipeline, not the kernel measurements taken inside that pipeline. The qReduMIS pipeline itself is considerably stronger and reaches the optimum on all nine instances.

Across our $43$-instance dataset, the seed pools hold between $2$ and $493$ distinct solutions (mean $57.0$, median $16$). On six instances (\texttt{L12\_seed2922}, \texttt{L14\_seed1590}, \texttt{L14\_seed2358}, \texttt{L14\_seed2480}, \texttt{L15\_seed3591}, and \texttt{L17\_seed424}) only one solution reaches the best MIS size, and the implementation pairs it with the best solution one atom smaller, so their pools of two include that solution. On $35$ of the $43$ instances that size is the certified optimum (Section~\ref{sec:validation}). Eighteen instances yielded fewer than $10$ unique solutions, $16$ produced between $10$ and $99$, and $9$ produced between $100$ and $493$. This regime ranging between approximately 10 and a few hundred solutions is characteristic of quantum annealers, which are known to exhibit biased sampling and do not find all ground-state configurations with equal probability for a given seed. This has been established both on superconducting annealing hardware~\cite{mandra:17,konz2019fairsampling} and, for Rydberg atom arrays of the kind used here, in Ref.~\cite{schuetz2025compilation}. That Houdayer cluster updates improve precisely this class of sampling bias was established extensively for spin-glass optimizers in earlier work~\cite{feeding2019}. Here we apply the same principle to independent sets returned by quantum hardware, with the added requirement that the hard independence constraint of the MIS be respected throughout. It is this limited and non-uniform coverage that motivates classical post-processing.

\subsection{Houdayer Cluster Updates}
\label{sec:houdayer_algo}

The Houdayer cluster update algorithm~\cite{houdayer2001} was originally developed for two-dimensional spin-glass systems and has since been extended to higher-dimensional binary systems in regimes (e.g., temperature or dimensionality) where clusters do not percolate \cite{zhu:15b} and thus are effective in resampling configurations of the system and improving thermalization when simulated via Monte Carlo methods. More recently, it has been applied to improve sampling from biased quantum and classical optimizers~\cite{feeding2019}. The core idea is to recombine pairs of solutions through non-local cluster exchanges, generating many new candidate configurations from a single pair in a rejection-free manner (i.e., not affecting the combined energy of the system).

Given two solution configurations $s_i$ and $s_j$, the algorithm first computes their symmetric difference $\Delta_{ij} = s_i \oplus s_j$, i.e., the set of vertices on which the two solutions disagree. The agreement backbone $s_i \cap s_j$ remains fixed throughout. The induced subgraph $G[\Delta_{ij}]$ is then decomposed into $c$ connected components, each representing an independent region of disagreement. Because these components are disconnected in $G[\Delta_{ij}]$, the vertex assignments within each cluster can be swapped independently without affecting any other cluster. Enumerating every binary exchange pattern, each specifying for each cluster whether to take vertices from $s_i$ or $s_j$, yields $2^c$ candidate solutions from a single pair. Every such candidate is independent by construction, because each cluster contributes vertices from one parent only, no edge of $G$ joins distinct components of $G[\Delta_{ij}]$, and a backbone vertex cannot be adjacent to a selected vertex of either parent. When $s_i$ and $s_j$ have equal size, a further consequence follows. Complementary candidates, obtained by exchanging every cluster, have total size $|s_i| + |s_j|$, so if that common size is the maximum independent set size neither candidate can exceed it and both must equal it. Every cluster of two maximum independent sets is therefore balanced between the two parents, and the exchange is size-preserving without any rejection step. If instead the parents share a suboptimal size, individual clusters may be imbalanced, and taking the larger side of such a cluster yields a strictly larger independent set. This is the mechanism behind the quality improvements of Section~\ref{sec:classical_improves_quantum}: for \texttt{L12\_seed829}, $17$ of the $2016$ seed pairs contain at least one imbalanced cluster, and for \texttt{L17\_seed46} all $6$ do, and each of those yields a valid solution at least one vertex larger than the seeds. For suboptimal parents a candidate can also be non-maximal, and the repair of Section~\ref{sec:repair} then adds vertices. All $2^c$ patterns are enumerated without a cap on $c$, whose largest observed value was $14$. 

Starting from the initial quantum seed pool $\mathcal{S}_0$ at the best-found size $k$, the algorithm iterates over the $\binom{|\mathcal{S}_0|}{2}$ pairs drawn from the seed pool, collecting newly discovered solutions of size $k$ into the output pool as they are found. Discovered solutions are not themselves re-paired: the pair list is fixed by $\mathcal{S}_0$, so the number of exchange operations is bounded a priori rather than growing with the pool, and within a pass every recovered configuration is a single recombination of two of its seeds. Solutions are stored as frozen sets of vertex indices for $\mathcal{O}(1)$ deduplication via hash lookup, so each configuration is counted only once regardless of how many pairs produce it. Candidates larger than $k$ are collected separately and, once the pass ends, the largest of them seed a new pass at that size, repeated until a pass finds nothing larger. These further passes are what enable classical post-processing to surpass the size of the seed pool on hard instances. The algorithm terminates with an expanded pool $\mathcal{S}_{\mathrm{final}}$ of optimum size $k_{\mathrm{final}} \geq k$, and the amplification factor $A = |\mathcal{S}_{\mathrm{final}}| / |\mathcal{S}_0|$ is recorded.

\subsection{MIS Repair}
\label{sec:repair}

Two operations in the pipeline can yield a set that is not a maximal independent set: raw hardware measurements, which may select adjacent atoms when the blockade is imperfect, and cluster exchange, which preserves independence (Section~\ref{sec:houdayer_algo}) but can leave a vertex addable. We handle both with a two-stage greedy procedure adapted from Ref.~\cite{qredumis2025}. Only its first stage is ever required for raw measurements, and only its second for exchange candidates. In the first stage, vertices are removed iteratively in order of decreasing conflict degree, the number of neighbors also in the current set, until no violations remain. This greedy ordering minimizes the number of removals needed to reach feasibility. In the second stage, vertices outside the current set are added greedily in order of increasing degree in the remaining graph, so that each addition leaves as many future candidates unblocked as possible. This extension continues until no further vertex can be added without creating a conflict, restoring maximality. The full procedure runs in $\mathcal{O}(|E|)$ time, with $E$ the number of edges of the graph, per candidate and in practice preserves the optimal set size for the large majority of repaired solutions. A small fraction of candidates repair to a smaller-than-optimal set and are discarded.

\subsection{Adaptive Stopping Criteria}
\label{sec:stopping}

Processing all $\binom{|\mathcal{S}|}{2}$ pairs is feasible when the solution pool is small, but becomes prohibitively expensive as the pool grows during exploration. We therefore implement an adaptive stopping strategy that selects among three modes at runtime. When the total pair count is below $10^4$, the algorithm runs exhaustively, guaranteeing complete pairwise coverage. For larger pools, it monitors the discovery rate over a sliding window of $500$ consecutive pairs and halts when the number of new solutions found within that window falls below $0.1$\% of the current pool size, indicating that the accessible manifold region has been saturated. A hard cutoff of $10^5$ pairs bounds the number of pair comparisons regardless of discovery dynamics, although this limit was never reached in our experiments. The largest number of pairs processed in any single pass was $16539$, and in any higher-level pass $6951$ (the second-level pass of \texttt{L17\_seed46}).

\subsection{Validation}
\label{sec:validation}

As a safeguard against implementation error, rather than as a test of the algorithm, all solutions in $\mathcal{S}_{\mathrm{final}}$ are validated post-hoc by verifying independence (no edge connects two selected vertices), maximality (no unselected vertex can be added without creating a conflict), and target cardinality (solution size equals $k_{\mathrm{final}}$). This last check confirms membership at the size the run reached, which coincides with the certified optimum only where the two agree. It is not an independent test of global optimality. The protocol was applied to all $15511$ solutions from the three instances where classical post-processing discovered a larger independent set than the quantum hardware, with a $100$\% pass rate. Throughout, the reference (true) maximum independent set size of each instance is taken from the reference values distributed with the data set of Ref.~\cite{qredumis2025,qredumis_github}. These values were computed with an exact sweeping-line solver, not with the qReduMIS pipeline~\cite{qredumis_github}. Our transfer-matrix dynamic programming over the lattice columns, exact for these geometries since all atoms occupy integer lattice sites and the blockade couples only adjacent columns, reproduces both the optimum size and the exact ground-state degeneracy for all $121$ reference instances that record them. Since the reference solver is also a sweep over the lattice, this agreement checks our implementation rather than constituting an independent certificate. It likewise reproduces every optimum quoted in this work, including all nine in Table~\ref{tab:improvement} and that of \texttt{L16\_seed4816} ($64$, with degeneracy $4480$). We use these reference sizes both to identify which instances quantum hardware failed to solve and to quantify the residual gap of the post-processed solutions (Table~\ref{tab:improvement}). Reference optima are available for $39$ of the $43$ instances. For the four Benchmark instances, which carry no reference entry, we recomputed the optima directly on the reconstructed unit-disk graphs, obtaining $|\mathrm{MIS}| = 46$, $47$, $45$, and $45$ for \texttt{HP125}, \texttt{HP1412}, \texttt{HP1435}, and \texttt{HP1478}, respectively. Two exact methods with different structure, branch-and-bound with reduction rules and the transfer-matrix dynamic programming above, agree on all four. The latter also supplies their exact degeneracies, which are used for the Benchmark family in Section~\ref{sec:coverage}. Comparing against these references, the quantum seed pool sat strictly below the true optimum on $8$ of the $43$ instances: six of the nine hard instances (all but \texttt{L12\_seed2922}, \texttt{L14\_seed2480}, and \texttt{L15\_seed3591}, whose single best seed already sits at the optimum), the Benchmark instance \texttt{HP1435} (size $44$ against a true optimum of $45$, consistent with its reported QAA success rate of $0$\%~\cite{qredumis2025}), and one standard instance, \texttt{L16\_seed4816} (size $63$ against $64$). On the remaining $35$ instances the seed pool already reached the certified optimum, so the configurations discovered at that size are optimal as well. This does not make the pool as a whole optimal: it also retains smaller configurations.

\section{Results}
 \label{sec:results}

\begin{figure*}
\includegraphics[width=\textwidth]{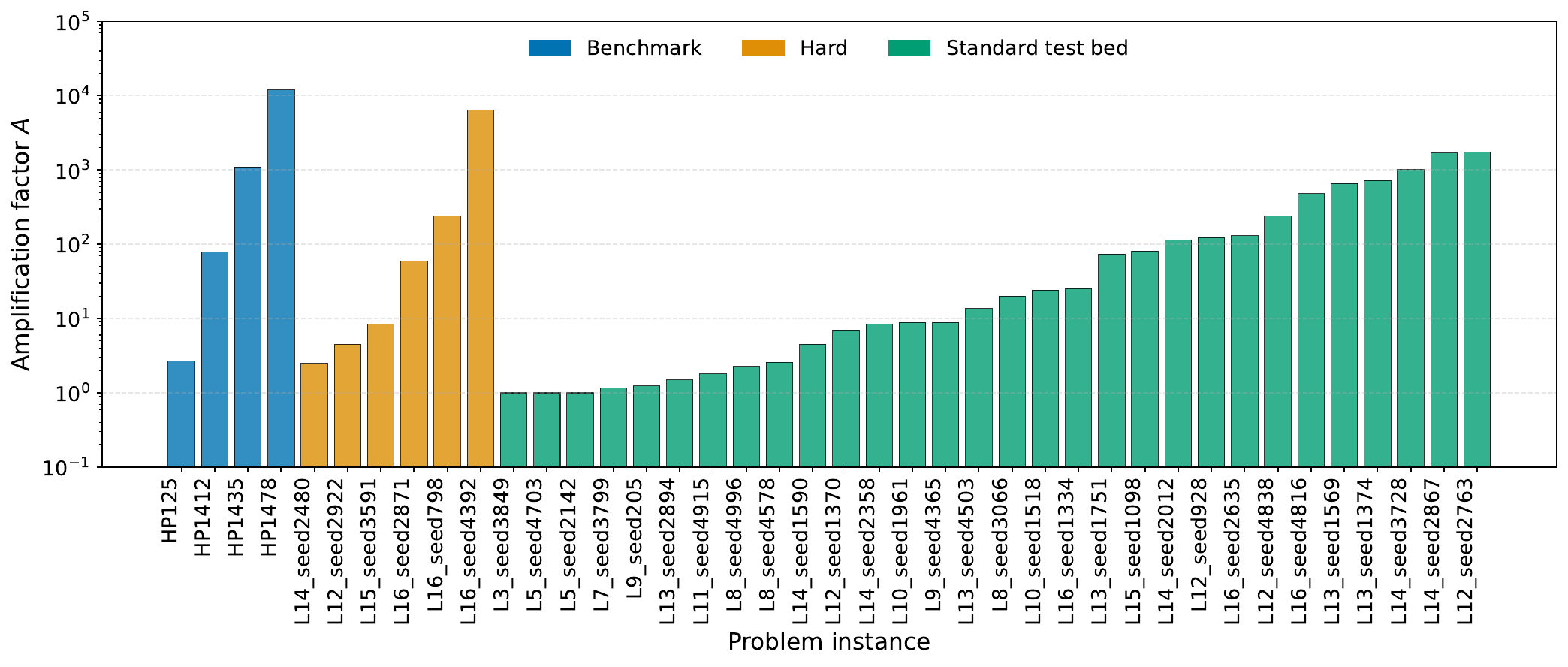}
\caption{\label{fig:amplification_overview} Amplification across the $40$ problem instances whose best independent-set size did not change under post-processing; the three instances where classical post-processing found a strictly larger independent set are shown separately in Table~\ref{tab:improvement}. The bar chart shows the amplification from initial quantum solutions to the final post-Houdayer solution pool on a base-10 log scale. Instances are grouped by type (Benchmark, hard, and standard). Across the full $43$-instance dataset, amplification factors range from $1.0$$\times$ to $11923$$\times$ with a median of $24.3$$\times$.}
\end{figure*}

\begin{figure}
\includegraphics[width=\columnwidth]{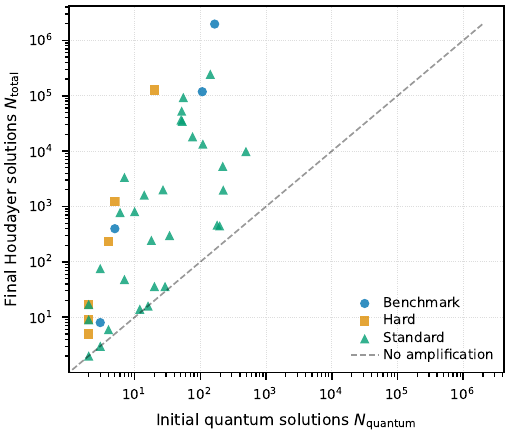}
\caption{\label{fig:initial_vs_final} Scatter plot showing the initial
quantum solution count $N_{\mathrm{quantum}}$ versus the final Houdayer
solution pool size on a base-10 log-log scale. Points above the diagonal
indicate amplification; the three points on the diagonal are instances that showed none. The $40$ instances shown exclude the three improvement
instances, which are treated separately (Section~\ref{sec:classical_improves_quantum}). Benchmark instances are represented by circles, hard instances by squares, and standard instances by triangles. The highest amplification
(\texttt{HP1478}) generates $1967339$ solution configurations from an
initial pool of $165$, i.e., a $11923$$\times$ amplification. Most instances show substantial amplification regardless of initial quantum sample size, demonstrating that solution count amplification is not simply proportional to quantum seed quantity.}
\end{figure}

\begin{figure}
\includegraphics[width=\columnwidth]{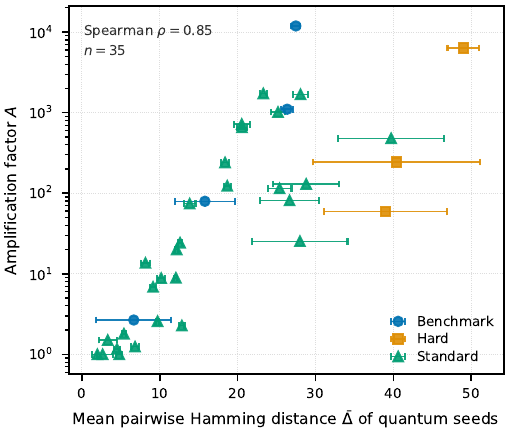}
\caption{\label{fig:initial_diversity} Initial solution diversity versus amplification factor. The mean pairwise Hamming distance $\bar{\Delta}$ (with error bars from a delete-one-seed jackknife, none for pools of two seeds) between the \emph{initial} quantum solutions versus amplification success. The seed pools are reconstructed from the raw hardware samples through the same repair procedure used by the algorithm (Section~\ref{sec:repair}), so $\bar{\Delta}$ characterises the pool the hardware actually supplied rather than the amplified manifold. Of the $43$ instances, $35$ are shown: the three improvement instances are treated separately (Section~\ref{sec:classical_improves_quantum}) and five further instances yield only a single distinct solution at the optimum, so they admit no pair. Benchmark instances are represented by circles, hard instances by squares, and standard instances by triangles. Because $\bar{\Delta}$ grows with manifold size, much of this association reflects $D_{\mathrm{MIS}}$ (Section~\ref{sec:diversity}).}
\end{figure}        

\begin{figure}
\centering
\includegraphics[width=\columnwidth]{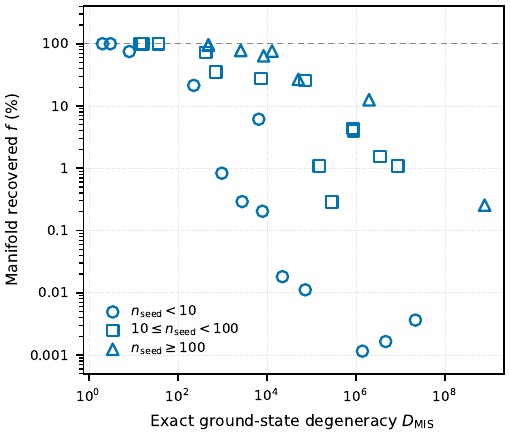}
\caption{\label{fig:manifold_coverage} Fraction of the exact ground-state manifold recovered, against the manifold's size. The reference data set records not only the optimum size but the exact degeneracy $D_{\mathrm{MIS}}$~\cite{qredumis2025,qredumis_github}, so coverage can be measured as $N_{\mathrm{total}}/D_{\mathrm{MIS}}$ rather than bounded. Only the $35$ instances whose seed pool already sits at the certified optimum are shown; the $8$ excluded are the shortfall instances of Section~\ref{sec:validation}, whose recovered solutions are not ground states, so coverage of the ground-state manifold is undefined for them. For the Benchmark instances, which have no reference entry, $D_{\mathrm{MIS}}$ is computed exactly by transfer-matrix dynamic programming over the lattice columns, a method validated against all $121$ reference instances whose atom positions are distributed with the dataset ($121$ of $121$ exact matches in both optimum size and degeneracy). Marker shape bins the size $n_{\mathrm{seed}}$ of the quantum seed pool, the second factor controlling the recovered fraction $f$. A joint fit over both predictors gives $f \propto D_{\mathrm{MIS}}^{-0.58 \pm 0.04} n_{\mathrm{seed}}^{1.40 \pm 0.12}$ with $R^2 = 0.89$: at fixed seed pool $f$ falls with manifold size, while at fixed manifold size it rises with the number of seeds.}
\end{figure}

\subsection{Overview of Amplification Results}
\label{sec:overview}

Figure~\ref{fig:amplification_overview} shows the amplification results for the $40$ instances whose best independent-set size did not change under post-processing, grouped by type. The three instances where post-processing found a strictly larger independent set are treated separately in Section~\ref{sec:classical_improves_quantum}. Across the full $43$-instance dataset, Houdayer cluster updates consistently expand the quantum-generated solution pools: amplification factors range from $1.0$$\times$ to $11923$$\times$, with a median of $24.3$$\times$ and a mean of $657.2$$\times$. The distribution is strongly right-skewed, driven by a handful of instances with unusually large degenerate manifolds. Amplification was not universal, however: three instances with more than one distinct quantum solution showed none at all (\texttt{L3\_seed3849}, \texttt{L5\_seed4703}, and \texttt{L5\_seed2142}, all at exactly $1.0$$\times$). These are the smallest in the test bed, at $7$, $20$, and $20$ atoms, and their seed pools already contained the complete ground-state manifold, so $1.0\times$ is the best attainable result rather than a failure (Section~\ref{sec:coverage}).

Figure~\ref{fig:initial_vs_final} shows the initial quantum solution counts against the final counts using Houdayer cluster updates. Most points lie above the diagonal, confirming amplification across all problem types and sizes. The three points on the diagonal are instances that showed no amplification. The spread along the vertical axis for fixed initial count illustrates that amplification is not simply proportional to the size of the quantum seed pool: instances with similar initial counts can differ by orders of magnitude in final count, reflecting differences in underlying manifold structure rather than sampling volume. Which of these two situations obtains is a property of the ground-state structure rather than of the pool size: Fig.~5 of Ref.~\cite{zhu:19} illustrates the distinction for two paradigmatic spin-glass instances, where a manifold made of many disconnected components in configuration space limits what any recombination move can reach, while a connected one does not.

The adaptive stopping criterion distributed efficiently across instances: $38$ of $43$ reached exhaustive pairwise exploration, while $5$ saturated before processing all pairs, indicating early convergence on large manifolds. The maximum number of pairs processed in any pass was $16539$, well below the $10^5$ pair hard cutoff, confirming that the saturation heuristic reliably identifies manifold boundaries without wasting computation time. We stress that ``exhaustive'' here refers to processing every pair \emph{within the given seed pool}. Such coverage still explores only the manifold region reachable from those seeds (Section~\ref{sec:limitations}), and does not imply discovery of the complete ground-state manifold or of the global optimum.

\subsection{Amplification Patterns Across Problem Types}
\label{sec:patterns}

Amplification behavior differs systematically across the three instance families, reflecting distinct relationships between quantum seed quality and manifold structure.

\textbf{Benchmark instances} with a hardness parameter $H \in \{1.478, 14.12, 125.5, 1435\}$, $137$ atoms each, span a wide range of amplification. We retain the labels of Ref.~\cite{qredumis2025}, which suppress the decimal point of $H$, so that \texttt{HP1478} denotes $H = 1.478$ while \texttt{HP1435} denotes $H = 1435$, and $H$ is quoted at each mention below. Amplification ranges from $2.7\times$ to $11923\times$. The low end is set by instance \texttt{HP125}, for which quantum hardware returned only $3$ seeds, leaving little structural diversity for cluster exchanges to exploit. The other three instances all amplify by at least $79$$\times$. Across all $43$ instances, amplification shows no significant correlation with $H$ (Spearman $\rho = -0.03$, $p = 0.86$), and coverage over the $35$ instances of Section~\ref{sec:coverage} shows none either ($\rho = -0.17$, $p = 0.33$). Within this family the pattern is nonmonotonic: for the easiest instance, \texttt{HP1478} ($H = 1.478$, QAA success rate $35.2\%$~\cite{qredumis2025}), the algorithm achieves the highest amplification of any instance in the dataset, expanding $165$ quantum solutions to $1967339$ configurations, yet the hardest instance, \texttt{HP1435} ($H = 1435$, QAA success rate close to $0$\%), still achieves $1104$$\times$ amplification from $107$ initial solutions, albeit at size $44$, one atom below its true optimum of $45$, so that the amplification enlarges a suboptimal manifold, while the intermediate-hardness \texttt{HP125} ($H = 125.5$) amplifies by only a factor $2.7$$\times$. Amplification is thus governed less by hardness per se than by the size of the accessible degenerate manifold, the structure of the ground-state manifold, and the size of the quantum seed pool. In this case \texttt{HP1478} combines a low-hardness, large-manifold landscape with a rich seed pool ($165$ solutions), whereas \texttt{HP125} is limited by its small initial seed pool ($3$ solutions). As such, having more solution candidates, even at less-than-optimal MIS sizes, might help in the amplification process.

\textbf{Hard instances} ($115$ -- $231$ atoms) span the widest range of amplification: the six instances whose size did not change (shown in Figure~\ref{fig:amplification_overview}) amplify by $2.5$$\times$ to $6366$$\times$ (\texttt{L16\_seed4392}, $20$ seeds to $127329$ configurations). They differ from the rest of the dataset in solution quality rather than quantity. As discussed in Section~\ref{sec:classical_improves_quantum}, the quantum hardware reached the true optimum on only three of the nine, each time with a single solution at that size, so this family accounts for six of the eight instances in the dataset where the hardware fell short (Section~\ref{sec:validation}). On three of them (treated separately) classical post-processing discovered strictly larger independent sets.

\textbf{Standard instances} ($7$ -- $205$ atoms) display the widest variation, from near-unity amplification on small instances to $1727$$\times$ on instance \texttt{L12\_seed2763}. The highest-amplification standard instances were \texttt{L12\_seed2763} ($142$ $\to$ $245173$ solutions), \texttt{L14\_seed2867} ($55$ $\to$ $92941$ solutions), and \texttt{L14\_seed3728} ($52$ $\to$ $52868$ solutions). These tend to be larger graphs with many atoms, where unit disk geometry naturally gives rise to large numbers of degenerate optimal configurations. One instance in this family, \texttt{L16\_seed4816}, is an exception to the pattern that standard test bed seeds are already optimal: its seven quantum seeds sit at size $63$, whereas the certified optimum is $64$ with degeneracy $4480$, so the $3359$ solutions Houdayer discovered from them are all one atom short. Cluster exchange found no size-$64$ configuration, consistent with the three suboptimal hard instances that likewise failed to improve.

Across all types, amplification correlates only moderately with problem size (Spearman $\rho = 0.55$, $p = 1 \times 10^{-4}$, between atom count and amplification factor across all $43$ instances), and more weakly with the number of quantum seeds ($\rho = 0.37$, $p = 0.02$). Neither is a strong predictor. The exact degeneracy is a far better one ($\rho = 0.86$ between $D_{\mathrm{MIS}}$ and $A$ over the $35$ instances whose seeds reach the optimum).

\subsection{Initial Diversity and Manifold Exploration}
\label{sec:diversity}

Figure~\ref{fig:initial_diversity} examines the relationship between the diversity of the quantum seed pool and the resulting amplification. We measure diversity as the mean pairwise Hamming distance $\bar{\Delta}$ between initial quantum solutions. For binary solution vectors the Hamming distance equals the size of the symmetric difference $|s_i \oplus s_j|$ introduced in Section~\ref{sec:houdayer_algo}, so we use the two terms interchangeably and denote the per-pair value $|\Delta_{ij}|$ and its mean $\bar{\Delta}$. Error bars come from a delete-one-seed jackknife, because pairs sharing a seed are not independent. Higher $\bar{\Delta}$ indicates that quantum hardware produced structurally varied seeds spanning a broader region of the solution space.

A strong positive correlation is visible between the initial mean Hamming distance $\bar{\Delta}$ and the amplification factor (Spearman $\rho = 0.85$ across the $35$ instances that admit at least one seed pair, five being excluded because only a single one of their returned solutions lies at the optimum size). Much of it reflects manifold size. Over the $30$ instances that also appear in Figure~\ref{fig:manifold_coverage}, $\bar{\Delta}$ correlates with $D_{\mathrm{MIS}}$ at $\rho = 0.96$, and at fixed $D_{\mathrm{MIS}}$ the partial rank correlation between $\bar{\Delta}$ and $A$ is $0.00$. Diversity does not predict coverage either ($\rho = -0.84$), because diverse seed pools come from large manifolds, where the recovered fraction is small.

Note that the relationship is not deterministic: some instances with moderate initial diversity achieve high amplification, and vice versa, indicating that graph topology and the specific positions of disagreement clusters also play important roles. A larger symmetric difference breaks into more exchangeable components and so yields more candidates per pair, but how many of them are new depends on the graph.

\subsection{Classical Post-Processing Improves on Quantum Hardware for Hard Instances}
\label{sec:classical_improves_quantum}

\begin{table}[t]
\centering
\caption{All nine hard instances, comparing the largest independent set in the seed pool (``Seed''), after Houdayer post-processing of those seeds (``Houdayer''), and after local search from the same seeds with the same number of candidate evaluations (``LS'', Section~\ref{sec:why_houdayer}), against the exact optimum (``True''), certified as described in Section~\ref{sec:validation}. The seed pool misses the true optimum on six instances, and on the other three holds a single solution at it. Houdayer cluster updates improve on the seeds on three, reaching the true optimum on \texttt{L12\_seed829} and coming within one atom of it on the two \texttt{L17} instances. The residual gap (``Gap'') is Houdayer minus True, i.e., the shortfall of the best post-processed solution against the true optimum (Section~\ref{sec:limitations}).}
\label{tab:improvement}
\begin{ruledtabular}
\begin{tabular}{lccccc}
Instance & Seed & Houdayer & LS & True & Gap \\
\hline
L12\_seed829 & 35 & 36 & 36 & 36 & $0$ \\
L12\_seed2922 & 35 & 35 & 35 & 35 & $0$ \\
L14\_seed2480 & 47 & 47 & 47 & 47 & $0$ \\
L15\_seed3591 & 54 & 54 & 54 & 54 & $0$ \\
L16\_seed2871 & 63 & 63 & 63 & 64 & $-1$ \\
L16\_seed4392 & 61 & 61 & 61 & 62 & $-1$ \\
L16\_seed798 & 60 & 60 & 60 & 61 & $-1$ \\
L17\_seed424 & 68 & 70 & 69 & 71 & $-1$ \\
L17\_seed46 & 67 & 70 & 70 & 71 & $-1$ \\\end{tabular}
\end{ruledtabular}
\end{table}

For the nine hard instances we can compare our results directly against the exact reference optima of Section~\ref{sec:validation}. This comparison establishes two facts. First, quantum hardware failed to reach the true optimum on six of the nine hard instances, falling short by $1$ to $4$ atoms (Table~\ref{tab:improvement}), and on the other three it returned only a single solution at the optimum. These six account for most of the eight shortfalls across the whole dataset (Section~\ref{sec:validation}). The difficulty of these instances is that unassisted quantum annealing does not sample the ground-state manifold fairly. Second, classical Houdayer post-processing of the quantum seeds discovered strictly larger configurations on three of these nine instances (\texttt{L12\_seed829}, \texttt{L17\_seed424}, \texttt{L17\_seed46}), by one atom for \texttt{L12\_seed829}, reaching its optimum, and by two and three atoms for the two \texttt{L17} instances. These three cases are gains in solution quality. On the other three suboptimal hard instances (\texttt{L16\_seed2871}, \texttt{L16\_seed4392}, and \texttt{L16\_seed798}) cluster exchanges yielded no improvement.

\textbf{\texttt{L12\_seed829}}: The seed pool held $64$ solutions at size $35$. Houdayer exploration of these seeds discovered solutions at size $36$, the certified optimum, and a second pass from them recovered all $18$ configurations of the ground-state manifold ($D_{\mathrm{MIS}} = 18$).

\textbf{\texttt{L17\_seed424}}: The seed pool held a single solution at size $68$, which the implementation pairs with the best solution at size $67$ (Section~\ref{sec:quantum_seeds}). The exchange of this one pair produced configurations at sizes $69$ and $70$, and a second pass found $2$ solutions at size $70$ ($+2$ atoms).

\textbf{\texttt{L17\_seed46}}: The seed pool held $4$ solutions at size $67$. Three passes reached size $69$ and then $70$, ending with $15491$ solutions at size $70$ ($+3$ atoms), all \emph{strictly better} than every seed.

The seeds are near-optimal configurations that happen to differ structurally in ways that, when recombined through cluster exchange, produce new configurations of size $k+1$ or larger. All $15511$ solutions at the improved sizes passed independence, maximality, and target-cardinality validation ($100$\% success rate). Except on \texttt{L12\_seed829}, these sizes remain below the certified optima (Table~\ref{tab:improvement}), so those configurations are maximal rather than maximum independent sets.

We emphasize that these improvements do not amount to recovery of the true optimum. Only \texttt{L12\_seed829} reaches its true optimum. The two \texttt{L17} instances end one atom short, and the three suboptimal \texttt{L16} instances do not improve (Table~\ref{tab:improvement}). These gaps are a direct consequence of the non-ergodicity discussed in Section~\ref{sec:limitations}. Classical recombination thus extracts strictly more from a fixed quantum sample than the hardware reports on its own, but it does not close the gap to the optimum in general: the qReduMIS pipeline of Ref.~\cite{qredumis2025} reaches all nine of these optima, and exact methods certify them (Section~\ref{sec:validation}). A local search from the same seeds reaches the same sizes on \texttt{L12\_seed829} and \texttt{L17\_seed46}, and one atom less on \texttt{L17\_seed424} at equal cost (Table~\ref{tab:improvement}, Section~\ref{sec:why_houdayer}).

That improvements occur only on hard instances is not coincidental. These are precisely the instances where quantum hardware leaves the largest gap to the optimum, and hence where recombination of near-optimal seeds has room to help. The size of the seed pool does not decide the outcome: \texttt{L17\_seed424} improved from a single seed at its best size, while \texttt{L16\_seed4392} did not improve from $20$, nor \texttt{HP1435} from $107$. What decides it is the mechanism of Section~\ref{sec:houdayer_algo}. A size improvement requires a cluster that is imbalanced between the two parents, and the three instances that improved are exactly the three whose seed pairs contain one ($17$ of $2016$ pairs for \texttt{L12\_seed829}, all $6$ for \texttt{L17\_seed46}, and the single mixed-size pair of \texttt{L17\_seed424}). None of the $5671$ pairs of \texttt{HP1435}, or of the pairs of the other four suboptimal instances, contains one.

\subsection{Manifold Coverage}
\label{sec:coverage}

\begin{figure}
\includegraphics[width=\columnwidth]{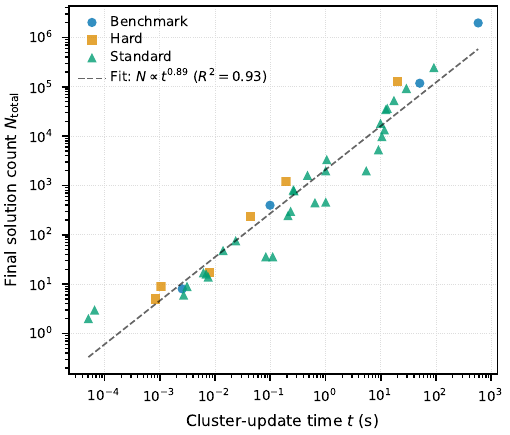}
\caption{\label{fig:computational_efficiency} Computational efficiency: cluster-update time versus final solution count. The one-off repair of the raw hardware samples is excluded here and is reported separately in the text, since its cost is set by the number of measurement shots rather than by the manifold recovered. All $43$ instances completed their cluster-update phase in under $10$ minutes on a standard laptop \cite{hardware}, with the largest manifold of approximately 2 million solutions requiring only $580$ seconds. The $40$ instances plotted exclude the three improvement instances, whose runs span two levels and are discussed separately (Section~\ref{sec:classical_improves_quantum}). The base-10 log-log scale reveals a near-linear scaling, $N_{\mathrm{total}} \propto t^{0.89}$ ($R^2 = 0.93$; dashed line), with substantial residual scatter. The hard cutoff of $10^5$ pair comparisons was never required.}
\end{figure}

Because the reference data set records the exact degeneracy $D_{\mathrm{MIS}}$ of the ground-state manifold alongside the optimum size, the coverage achieved by the method can be measured directly rather than merely bounded. Figure~\ref{fig:manifold_coverage} shows this for the $35$ instances whose quantum seeds already sit at the certified optimum, the only instances for which the question is well posed. The $8$ excluded are the shortfall instances of Section~\ref{sec:validation}.

Coverage spans five orders of magnitude, from $0.001$\% to $100$\%, with a median of $21.4$\%. At the top of that range the non-ergodicity of the method is not a limitation at all: on six instances the recovered pool is provably the \emph{complete} ground-state manifold (\texttt{L3\_seed3849} $3/3$, \texttt{L5\_seed4703} $2/2$, \texttt{L7\_seed3799} $14/14$, \texttt{L5\_seed2142} $16/16$, \texttt{L9\_seed205} $36/36$, and \texttt{L11\_seed4915} $36/36$), and two further instances exceed $94$\% (\texttt{L8\_seed4578} $463/479$, \texttt{L8\_seed4996} $447/474$). For these, cluster exchange from the quantum seeds enumerates the manifold exhaustively rather than sampling it, as certified against the exact degeneracies of Section~\ref{sec:validation}. All eight lie at the small end of the test bed, spanning $7$ to $97$ atoms against a median of $115$ across the instances plotted, which is the regime in which a few hundred measurements can cover a manifold outright.

Coverage falls steeply as the manifold grows (Spearman $\rho = -0.79$, $p = 2 \times 10^{-8}$): the median is $100$\% for the seven instances with $D_{\mathrm{MIS}} < 10^2$ but only $1.1$\% for the sixteen with $D_{\mathrm{MIS}} \geq 10^4$. The largest manifold belongs to \texttt{HP1478}, the instance with the highest amplification in the dataset: its $1967339$ recovered solutions, a $11923\times$ expansion of the quantum pool, represent $0.26$\% of a ground-state manifold of $D_{\mathrm{MIS}} = 7.7 \times 10^8$. Manifold size is not the only factor, however, and the two must be separated before the trend can be interpreted. Coverage also depends on how many seeds the hardware supplied, and across these $35$ instances the seed pool spans $2$ to $493$ solutions. Writing $f = N_{\mathrm{total}} / D_{\mathrm{MIS}}$ for the recovered fraction of the manifold, a joint fit of both predictors gives 
\begin{equation}
f \propto D_{\mathrm{MIS}}^{-0.58 \pm 0.04} \, n_{\mathrm{seed}}^{1.40 \pm 0.12}
\end{equation}
with $R^2 = 0.89$, and the two effects are close to independent in this dataset (Spearman $\rho = +0.17$ between $n_{\mathrm{seed}}$ and $D_{\mathrm{MIS}}$, $p = 0.3$). Controlling for seed count therefore \emph{strengthens} the degeneracy trend rather than explaining it away: the partial rank correlation between coverage and $D_{\mathrm{MIS}}$ at fixed $n_{\mathrm{seed}}$ is $-0.94$, against $-0.79$ for the raw pairwise value, while the partial correlation with seed count at fixed degeneracy is $+0.85$. The marker shapes in Figure~\ref{fig:manifold_coverage} show both factors acting together: binning by seed pool as in the seed statistics above, median coverage rises from $0.3$\% for the thirteen instances with fewer than $10$ seeds, to $26.6$\% for the fourteen with $10$ to $99$, to $70.2$\% for the eight with $100$ or more. This is the expected behaviour of a method working from a bounded seed budget, and it sets the trade-off. The absolute number of solutions recovered grows with manifold size (Spearman $\rho = +0.66$ between $D_{\mathrm{MIS}}$ and final count), while the \emph{fraction} recovered falls. Applications needing a fixed number of diverse solutions are therefore well served across the whole range, whereas applications needing a guaranteed-complete manifold are served only when the manifold is small enough to enumerate from the seeds available. Of the two factors, only the seed count is under experimental control, since $D_{\mathrm{MIS}}$ is a property of the instance.

\subsection{Computational Efficiency}
\label{sec:efficiency}

Figure~\ref{fig:computational_efficiency} shows cluster-update time versus final solution count for the $40$ instances whose size did not improve. All runs completed on a standard laptop \cite{hardware}, and the entire dataset, $2451$ quantum seeds expanded to $2749146$ validated configurations, required $1073$ seconds of computation end to end, i.e., $18$ minutes on a single laptop core. Almost all of it is the cluster-update loop: $870$ seconds across the $43$ first-level passes and $200$ seconds for the higher-level passes of the three multilevel runs. The one-off repair of every post-selected hardware sample to a maximal independent set (Section~\ref{sec:repair}) takes $2$ seconds for all $18759$ samples. Both stages are recovery costs. The post-hoc validation of Section~\ref{sec:validation} and the coverage certification of Section~\ref{sec:coverage} are analysis costs and are excluded. Of those configurations, $2485881$ are certified optimal against the reference values. The remaining $263265$, or $9.6$\%, belong to the $8$ instances whose seed pool sat below the certified optimum (Section~\ref{sec:validation}) and are validated at their best-found size rather than at the true optimum. The post-processing cost is small compared with the hardware time already expended: at a cycle time of about $0.1$\,s~\cite{qredumis2025}, a $1000$-shot run takes about $100$\,s of device time per instance. Per family:

\begin{itemize}
\item \textbf{Benchmark instances}: from under $1$ second (\texttt{HP125}, \texttt{HP1412}, with only $3$ and $15$ pairs) up to $580$ seconds for instance \texttt{HP1478} ($13530$ pairs), spanning nearly four orders of magnitude in runtime.
\item \textbf{Hard instances}: from under $1$ second up to $20$ seconds (\texttt{L16\_seed4392}, $190$ pairs), all reaching exhaustive pairwise exploration of their first-level seed pools. The two higher-level passes of \texttt{L17\_seed46} halted on saturation instead and took $200$ seconds together.
\item \textbf{Standard instances}: from under $1$ second to approximately $90$ seconds, depending on manifold size.
\end{itemize}

Figure~\ref{fig:computational_efficiency} plots cluster-update time rather than end-to-end time, because it is the exploration loop whose cost scales with the manifold being recovered. On that axis the log-log plot reveals a near-linear scaling of solution count with runtime, $N_{\mathrm{total}} \propto t^{0.89}$ ($R^2 = 0.93$), the exponent slightly below unity indicating modestly diminishing returns on additional computation. Substantial scatter remains about this trend: instances with comparable run times can differ by more than an order of magnitude in final count, because the per-pair cost (set by cluster count and repair complexity) varies by more than three orders of magnitude across the dataset, while the number of pairs required depends on seed-pool size. 

The hard cutoff of $10^5$ pair comparisons was never invoked, since the pass with the most pairs processed required only $16539$. Combined with the modest hardware requirements, these results suggest Houdayer post-processing is a practical and lightweight component of any quantum optimization pipeline, as can also be seen in Fig.~\ref{fig:computational_efficiency} where the scaling is close to linear.

\section{Discussion}
\label{sec:discussion}

\subsection{Why Houdayer Cluster Updates Work for the MIS Problem}
\label{sec:why_houdayer}

The effectiveness of Houdayer cluster updates for MIS manifold exploration rests on a combination of structural and algorithmic properties that make it well-matched to the geometry of degenerate combinatorial optimization problems, especially for unit-disk quasi-two-dimensional graphs. 

The key mechanism is nonlocal, rejection-free recombination. Local search moves such as single-vertex swaps~\cite{andrade2012} change one site at a time. Houdayer updates instead identify entire connected components of disagreement between two solutions and swap them en bloc, enabling large-scale structural rearrangements in a single operation. Because the agreement backbone is preserved and every candidate is independent by construction (Section~\ref{sec:houdayer_algo}), cluster exchanges are accepted with certainty. There is no energy barrier to overcome and no rejection step. To quantify the difference, we ran a local search from the same seed pools that moves a selected vertex to a free site with exactly one selected neighbor, keeping the set size, and whenever such a move frees a vertex adds it and continues from the larger set, with the same number of candidate evaluations as the cluster updates used on each instance. On the $30$ instances with optimal seeds the two recover comparable numbers of configurations (local search more on $14$, cluster updates more on $10$, median ratio $1.00$), but cluster updates recover more on $7$ of the $8$ largest manifolds, $1967339$ against $911142$ for \texttt{HP1478}, where one pair yields many configurations at once. From suboptimal seeds, the two methods improve the same three instances and neither improves the other five. Local search reaches the same sizes on \texttt{L12\_seed829} and \texttt{L17\_seed46} and one atom less on \texttt{L17\_seed424}, where it needs about $3 \times 10^4$ evaluations to reach the size-$70$ configurations that cluster updates find with $66$ (Table~\ref{tab:improvement}). All configurations from both methods were validated as described in Section~\ref{sec:validation}.

For the MIS on unit disk graphs, the sparse and geometrically constrained edge structure causes the induced subgraph $G[\Delta_{ij}]$ to break into several components, so a single pair generates $2^c$ candidates: over the $135202$ seed pairs of the $30$ instances with optimal seeds (at most $20000$ per instance), the median is $c = 3$ and the maximum $c = 13$. The components are not restricted to the two-vertex form drawn in Figure~\ref{fig:workflow}: $57$\% contain one vertex from each parent and the remainder are larger, up to $44$ vertices. The fan-out is unaffected by component size, since each component contributes one binary choice however many vertices it holds. Between maximum independent sets every component is balanced, and none of these pairs contains an imbalanced one. Between the suboptimal seeds of the other $7$ instances with at least two of them, $0.16$\% of components are imbalanced, all but one of odd size, and these are the source of the size improvements in Section~\ref{sec:classical_improves_quantum}. The exponential fan-out per pair is what drives the rapid manifold expansion observed across all instance types.

The complementarity with quantum sampling is also important. Quantum annealing explores the solution space through coherent tunneling and tends to cluster samples near local energy minima. Houdayer updates operate orthogonally: they systematically recombine structurally distinct solutions to reach regions of the ground-state manifold that neither single-solution perturbation nor further quantum sampling would easily access. The two methods are thus complementary. Quantum provides high-quality seeds, and classical cluster updates provide systematic coverage.

\subsection{Two Regimes: Amplification and Quality Improvement}
\label{sec:two_regimes}

Our results reveal two qualitatively distinct operating regimes for hybrid Houdayer cluster-update post-processing, determined by the relationship between quantum hardware capability and ground-state manifold structure.

In the \textbf{large manifold regime}, quantum hardware reaches the true ground state (verified against the reference optima, Section~\ref{sec:validation}) and produces an initial pool of diverse optimal solutions. Classical post-processing then systematically expands this pool by recombining seeds to reach structurally distant manifold regions. This regime dominates for the standard test bed and for those benchmark instances that combine a large degenerate manifold with a diverse seed pool, where amplification factors can reach $10^3$ -- $10^4$ times the original pool.

Amplification and solution quality are, however, largely decoupled, and the clearest illustration is instance \texttt{HP1435}. It is the hardest Benchmark instance ($H = 1435$, QAA success rate $0$\%~\cite{qredumis2025}), its quantum seeds sit one atom below the true optimum of $45$, and Houdayer post-processing nonetheless amplified them $1104$$\times$ into $118153$ configurations, all at the suboptimal size $44$. A large, diverse seed pool therefore guarantees a large final pool but says nothing about whether that pool sits at the optimum. Amplification quantity scales with the accessible manifold size and seed diversity rather than with hardness, and independently of whether the seeds are optimal: larger degenerate spaces offer more recombination opportunities, and high initial diversity (Figure~\ref{fig:initial_diversity}) seeds broader exploration.

In the \textbf{difficult landscape regime}, quantum hardware does not reliably find the true ground state. The initial solution pool sits at a suboptimal energy level, and the Houdayer algorithm by operating on these seeds can construct recombinations that repair into higher-quality configurations. This regime characterizes the three hard instances where classical post-processing discovered strictly larger independent sets. Amplification in this regime is variable (from $10\times$ for instance \texttt{L17\_seed424} up to $1139\times$ for instance \texttt{L17\_seed46}). Both figures refer to the first-level pass, which expands the quantum seed pool at the seed size and is where the larger configurations are first discovered. The subsequent pass seeded by those configurations amplifies a different, smaller starting set, so ratios across the two levels are not comparable. The spread reflects how many configurations exist at the improved size, but the qualitative outcome, improvement in solution quality, is of greater practical interest than the raw count. It demonstrates that classical recombination can partially compensate for systematic biases in quantum sampling, and in one case close the gap to the true optimum.

Note that both regimes are not mutually exclusive, and this suggests a strategy we have not tested here. For the pure Houdayer update the combined energy of the two-replica system is conserved, so pairing sub-optimal solutions should trade optimality between the two members of a pair: as one improves the other degrades, and retaining only the better member would drive a pool of sub-optimal solutions towards optimality. We note two caveats. The MIS repair step (Section~\ref{sec:repair}) changes set sizes and therefore breaks the exact energy conservation the argument relies on, and no experiment in this work seeds the algorithm with a deliberately sub-optimal pool. We therefore offer this as a direction for future work rather than an established mechanism.

Classical post-processing thus provides a computationally cheap step that can partially mitigate poor quantum hardware performance. 

\subsection{Limitations, Non-Ergodicity, and Future Directions}
\label{sec:limitations}

There are, however, limitations to cluster-update-based post-processing approaches. First, the Houdayer algorithm is fundamentally non-ergodic: it can only reach configurations obtainable by a single cluster exchange between two seeds. If the quantum seeds are all confined to a single pocket of the solution space, the algorithm will map only that disconnected pocket and leave other regions entirely unexplored. Coverage rises with the size of the seed pool and falls with the size of the manifold (Section~\ref{sec:coverage}). Increasing the number of quantum shots or diversifying the annealing schedule, for which systematic design methods now exist~\cite{finzgar2024}, would enlarge and diversify the seed pool and, in turn, improve manifold coverage. A further possibility for future work is to pool seeds from solvers with different sampling biases, for example quantum measurements together with a cheap classical heuristic, so that the sources populate complementary regions of the manifold. Alternatively, including less optimal solutions can allow the algorithm to climb over small barriers in the energy landscape by successively applying cluster updates. Second, we have not run the converse control of seeding the algorithm with cheap classical near-optima, either in place of the quantum seeds or mixed with them, so the marginal value of the quantum seeds specifically, as opposed to any sufficiently diverse near-optimal pool, is not isolated here. Ref.~\cite{feeding2019} addresses the bias-correction question for classical optimizers but not this comparison. Third, the approach is heuristic, i.e., there is no \emph{a priori} guarantee that all optima are found. Where a reference degeneracy is available this can be checked rather than assumed, and Section~\ref{sec:coverage} does so: coverage of the ground-state manifold ranges from $0.001$\% to $100$\%, and on six instances the recovered pool is provably complete. Completeness is therefore attainable, but only for manifolds small enough that a fixed seed budget can enumerate them. For applications that require solution diversity rather than completeness, a finite number of uncorrelated solutions is often sufficient.  Fourth, the exploration scales as $\mathcal{O}(|\mathcal{S}_0|^2)$ in the size of the \emph{seed} pool, which becomes costly for very large seed sets. However, our saturation-based stopping criterion mitigates this in practice.  Finally, all instances in this study are unit disk graphs arising from Rydberg atom array geometries. The effectiveness of Houdayer updates on graphs where clusters percolate (e.g., power-law distribution of nodes as in scale-free graphs or higher dimensions) degrades rapidly. However, some improvement might be achievable even in these cases for sections of the ground-state manifold.

Houdayer cluster updates are not specific to MIS. The core mechanism of symmetric difference decomposition into independent clusters, rejection-free exchange applies to any combinatorial problem where solutions can be represented as binary assignments and pairs of solutions generate a well-defined overlap structure. It carries over without a new repair step to maximum clique and minimum vertex cover, through their equivalence to the MIS problem, and to weighted MIS, where the exchange still preserves independence and changes the weight by the difference between the two sides of each cluster. Other candidates, such as constraint satisfaction problems with degenerate solution spaces, would need a repair step adapted to their feasibility structure.

\section{Conclusion}
\label{sec:conclusion}

We have measured how much ground-state information a neutral atom MIS measurement record contains beyond the solutions the device reports. Across $43$ instances from QuEra's Aquila Rydberg device, classical post-processing via Houdayer cluster updates recovers between $1.0\times$ and $11923\times$ as many distinct configurations at the certified-or-best-found size (median $24.3\times$, and $8.9\times$ over the $35$ instances whose seeds were already optimal), turning pools of tens to hundreds of configurations into collections with a median of $463$ and as many as nearly two million. The entire dataset required $18$ minutes of computation on a single laptop core, measured end to end, compared with about $100$\,s of device time for each $1000$-shot hardware run.

Two features of this data set make the measurement possible. First, exact ground-state degeneracies are available for these instances, so the recovered fraction is a number rather than a bound: a median of $21$\% of the true manifold, and on six instances provably all of it. The non-ergodicity of the method is thus quantified rather than disclaimed, and its practical boundary identified, namely that completeness is reachable when the manifold is small enough for a fixed seed budget to enumerate, and that coverage falls steeply as the manifold grows. Second, the same exact references reveal that the quantum seed pool sat below the true optimum on $8$ of the $43$ instances. Amplification and solution quality are therefore separate axes: a run can be amplified a thousandfold and still describe a suboptimal manifold, and only an exact reference distinguishes the two cases.

For three hard instances, classical post-processing discovered independent sets strictly larger than any in the seed pool, and on one of them, \texttt{L12\_seed829}, it reached the certified optimum and recovered its entire $18$-configuration manifold. All $15511$ improved solutions passed independence, maximality, and target-cardinality validation. Because a local search from the same seeds reaches the same sizes on two of the three, the improvements show what recombination can extract from near-optimal seeds rather than an advantage specific to cluster updates.

The practical consequence is that solution counts and diversity metrics reported directly from quantum hardware understate what a run has determined, by a median factor of $24$ on this dataset and by up to four orders of magnitude in the extreme case. For any application whose figure of merit is the number or diversity of optimal configurations rather than a single best solution, the quantity worth reporting is the post-processed count, since the recovery costs seconds to minutes of laptop time per instance. Because the cluster exchange mechanism is not specific to MIS, the same measurement can be made for any problem whose solutions admit a binary encoding and a well-defined overlap structure, which is where we expect this accounting to matter next.

\begin{acknowledgments}

AJO would like to thank Graham Enos for his careful review of this manuscript. HGK would like to thank Jon Machta for insightful conversations on cluster updates and thanks Mark Lundgaard, Rasmus Munk, Nicolai Nørregaard, and Eric Vildgaard for inspiring the work. MJAS would like to thank Sri Elaprolu for his leadership and support. This work received no external funding.

The authors acknowledge the use of AI tools to assist with code generation, data analysis, and drafting and editing parts of the manuscript text. The manuscript was written and reviewed in depth by the authors, who take full responsibility for its content.

\end{acknowledgments}

\par\null\vspace{-\baselineskip}
\section*{Data and Code Availability}

The quantum-hardware measurement records, problem instances, and reference values analyzed in this work are those distributed with Ref.~\cite{qredumis2025} and are publicly available at Ref.~\cite{qredumis_github}.

\bibliography{refs}

\end{document}